\documentclass[iop]{emulateapj}

\shorttitle{DAZ white dwarf NLTT~25792}
\shortauthors{Vennes \& Kawka}

\def\kms{km~s$^{-1}$}
\def\teff{$T_{\rm eff}$}
\begin{document}

\title{The polluted atmosphere of the white dwarf NLTT~25792 and the diversity of circumstellar environments $^*$}

\altaffiltext{*}{Based on observations collected at the European Organisation for
Astronomical Research in the Southern Hemisphere, Chile under programme
ID 091.D-0267.}

\author{S. Vennes and A. Kawka}
\affil{Astronomick\'y \'ustav, Akademie v\v{e}d \v{C}esk\'e republiky, Fri\v{c}ova 298, CZ-251 65 Ond\v{r}ejov, Czech Republic}

\begin{abstract}
We present an analysis of X-Shooter spectra of the polluted, hydrogen-rich white dwarf NLTT~25792. The spectra show
strong lines of calcium (Ca~H\&K, near-infrared calcium triplet, and \ion{Ca}{1}$\lambda$4226) 
and numerous lines of iron along with magnesium and aluminum lines from which we draw the abundance pattern.
Moreover, the photospheric Ca~H\&K lines are possibly blended with a circumstellar component shifted by $-20$ \kms\ relative to the 
photosphere. 
A comparison with a sample of four white dwarfs with similar parameters show considerable variations in their abundance patterns,
particularly in the calcium to magnesium abundance ratio that varies by a factor of five within this sample.
The observed variations, even after accounting for diffusion effects, imply similar variations in the putative 
accretion source.
Also, we find that silicon and sodium are significantly underabundant in the atmosphere of NLTT~25792,
a fact that may offer some clues on the nature of the accretion source.
\end{abstract}


\section{Introduction}

The presence of heavy elements in the atmosphere of many cool white dwarfs is attributed
to circumstellar debris that are continuously or intermittently accreted onto the white dwarf surface \citep[see][]{zuc2003}.
Questions remain as to the phase and composition of the accreted material and its effect on the white dwarf abundance pattern.
\citet{zuc2003} and \citet{koe2005} estimate that approximately 25\% of cool, H-rich white dwarfs show the presence of metal lines
in their spectra, while \citet{zuc2010} found that a similar fraction of He-rich white dwarfs show metal lines.
\citet{kil2006} explored the link between heavy element pollution in white dwarfs and the presence of
a warm, near-infrared (IR) debris disk feeding the photosphere as in the case of the proto-typical dusty,
polluted white dwarf G29-38 \citep{jur2003}.
However, \citet{far2009} reported 
that only 21\% of polluted white dwarfs have mid-IR excess consistent with a circumstellar disk 
and they also noted
an increased likelihood of a mid-IR excess for objects with a higher calcium abundance
($\log{\rm Ca/H(e)} \ga -8.0$). Actually,
\citet{far2010} observed several polluted H-rich (DAZ) and He-rich (DZ)
white dwarfs with {\it Spitzer} and found that the debris disks have varying
thickness and, therefore, some of these may be narrow
enough to escape detection. Moreover,
\citet{jur2008} shows that polluted white dwarfs
that do not exhibit an IR excess may still be experiencing
gas accretion that originated from the tidal destruction of a large number
of small asteroids. Indeed, gaseous disks around white dwarfs were detected via 
near-IR calcium triplet emission \citep{gae2006,gae2007}.

Various accretion scenarios or many types of sources may be involved. For example, the immediate environment of white dwarf stars
is likely composed of remnant bodies that survived post-asymptotic giant branch evolution. These bodies may be  
low-mass stellar \citep[see, e.g.,][]{kaw2008} or sub-stellar companions \citep[see, e.g.,][]{ste2013}, or disrupted planetary systems, i.e., objects that are otherwise quite common. The formation of
a debris disk could mix material and produce an abundance pattern averaged over several constituents. Consequently,
the observed abundance would be representative of the stellar neighborhood. However, single
body accretion could deliver a greater diversity of abundance patterns \citep[see, e.g.,][]{zuc2011}.

The accretion history and diffusion time-scales mingle in a complex manner \citep{dup1992}. \citet{koe2009} considered three
possible sequences of events: One of continuous accretion and built-up toward diffusive equilibrium, one
at diffusive equilibrium, and, after extinction of the accretion source, one of decline in photospheric abundances. 
Therefore, the observed abundance pattern, i.e., the photospheric abundance ratios, may differ considerably from the source pattern,
and a case-by-case study of DAZ white dwarfs may yet reveal considerable abundance variations.

On last count \citep[see, e.g.,][]{zuc2003,koe2005}, cool DAZ white dwarfs ($T_{\rm eff} \lesssim 8000$ K) are outnumbered at least 1:3
by the DZ white dwarfs \citep[see, e.g.,][]{duf2007}.
Interestingly, the number of identified DAZ white dwarfs in the local sample of \citet{gia2012}
is comparable to that of DZ white dwarfs (11 versus 13 stars),
although the DAZ count corresponds to a much smaller fraction of all H-rich white dwarfs 
than that of their He-rich counterparts, i.e., one tenth versus one third, respectively.
Clearly, many polluted H-rich white dwarfs remain to be recognized as such in the local sample.
\citet{koe2009} cites difficulties in detecting weak metal lines in the opaque, neutral hydrogen environment of cool DA white dwarfs. 
However, recent
spectroscopic surveys of local white dwarf candidates \citep[see, e.g.,][]{kaw2006,kaw2012a} have uncovered several new cool DAZ white dwarfs 
\citep{kaw2011,kaw2012b}.

In this context, we present an analysis of the polluted, otherwise hydrogen-rich white dwarf NLTT~25792 from the revised
New Luyten Two-Tenth catalog \citep{sal2003}.
\citet{kaw2011b} and \citet{gia2011} reported the detection of a strong Ca~K line but without other identifiable elements.
This object is also known in the Edinburgh-Cape catalog as EC~10542$-$2236 \citep{kil1997} and in the Luyten-Palomar catalog as LP~849-31.
NLTT~25792 is located above the Galactic plane ($l=271.1687,\,b=+32.8146$) and
toward a relatively tenuous line of sight in the interstellar medium \citep[$E_{B-V}=0.0595$,][]{sch1998}.
The proper motion quoted in the rNLTT catalog 
is $(\mu_\alpha\,\cos{\delta}, \mu_\delta)=(-72, 293)$ mas yr$^{-1}$.
Section 2 details our recent observations of this object using the X-shooter spectrograph at the European Southern Observatory,
and in Section 3 we present an analysis of this and other related polluted white dwarfs.
We conclude in Section 4.

\section{Observations}

\begin{deluxetable}{llc}
\tablecaption{Photometry\label{tbl_phot}}
\tablewidth{0pt}
\tablehead{
\colhead{Survey and band\tablenotemark{a}} & \colhead{$\lambda$ effective} & \colhead{$m$}   }
\startdata
{\it GALEX} NUV  &   2271 \AA     &   $17.78\pm0.03$ \\
SDSS $u$        &   3551 \AA     &   $16.540\pm0.007$    \\
SDSS $g$        &   4686 \AA     &   $16.079\pm0.004$    \\
SDSS $r$        &   6116 \AA     &   $15.974\pm0.004$    \\
SDSS $i$        &   7481 \AA     &   $15.984\pm0.005$    \\
SDSS $z$        &   8931 \AA     &   $16.062\pm0.007$    \\
2MASS $J$        &   1.235 $\mu$m  &  $15.521\pm0.050$   \\
2MASS $H$        &   1.662 $\mu$m  &  $15.401\pm0.111$  \\
2MASS $K$        &   2.159 $\mu$m  &  $15.943\pm0.255$   \\
{\it WISE} $W1$  &   3.353 $\mu$m  &  $15.268\pm0.046$   \\
{\it WISE} $W2$  &   4.603 $\mu$m  &  $15.051\pm0.103$       
\enddata
\tablenotetext{a}{{\it GALEX} GR6/GR7 photometry obtained at galex.stsci.edu/GalexView/;
SDSS Photometric Catalog, Release 7 \citep{aba2009};
2MASS photometry \citep{skr2006}; 
({\it WISE}) photometry \citep{cut2012}.
}
\end{deluxetable}

We obtained two consecutive sets of echelle spectra using the X-shooter spectrograph \citep{ver2011}
attached to the UT2 (Kueyen) at Paranal Observatory on 
UT 2013 May 10. 
The observations were conducted in clear sky conditions at an average airmass of 1.24 in the first set and 1.60 in the second. 
The seeing conditions were 
1.20 arcsec on average ($\sigma=0.11$ arcsec) during the first observation and 1.36 arcsec
($\sigma=0.13$ arcsec) during the second.
The spectra were obtained at the parallactic angle and in the ``stare'' mode, 
with the slit width set to 0.5, 0.9 
and 0.6 arcsec for the UVB, VIS and NIR arms, respectively. 
This arrangement delivered 
a resolution of $R=\lambda/\Delta\lambda = 9100$, 8800 and 6200 for the UVB, VIS and NIR arms, respectively, for 
nominal wavelength ranges of 2940-5930 \AA\ with UVB, 5250-10490 \AA\ with VIS,
and 0.98-2.48 $\mu$m with NIR. For each set,
the exposure times were 2940 and 3000 s for the UVB and VIS arms, respectively. 
For the NIR arm we obtained five exposures of 600 s each. An analysis of the NIR observations will be presented elsewhere.

We reduced the observations using the X-shooter reduction pipeline under the ESO Recipe Flexible Execution
Workbench (Reflex). Details of the X-shooter pipeline and Reflex are available in the ESO documents
VLT-MAN-ESO-14650-4840 (issue 12.0)  and VLT-MAN-ESO-19540-5899 (issue 2.0). The extracted spectra were resampled with 0.2 \AA\ bins, corresponding
to half of a resolution element in the UVB spectra, and one third of a resolution element in the VIS spectra.
The signal-to-noise ratio (SNR) achieved in the co-added UVB spectrum at $\lambda=$3100 \AA\ was SNR$\sim 10$ per bin 
thereby setting the
lowest usable wavelength. The measured SNR reached $\sim$30 per bin at 3500 \AA, $\sim$80 at 3900 and 4200 \AA, and $\sim$90 at 5000 \AA.
The measured SNR reached $\sim$54 per bin in the co-added VIS spectrum at 5900 \AA, $\sim$70 at 6600 \AA, and $\sim$90 at 8600 \AA.

Table~\ref{tbl_phot} lists available photometric measurements from the Galaxy Evolution Explorer ({\it GALEX}) sky survey,
the Sloan Digital Sky Survey (SDSS), the Two Micron All Sky Survey (2MASS), and the Wide-field Infrared Survey Explorer 
({\it WISE}).

\subsection{Comparison Sample}

Also, we obtained a series of HIRES spectra ($R=25000$ to 40000) for a set of closely
related DAZ white dwarfs: WD~0208$+$396 (G74-7), WD~0354$+$463, WD~1257$+$278, and WD~1455$+$298 \citep{zuc2003,zuc2011}.
\citet{zuc2003} published the spectra for WD~0208$+$396 (with the Keck Observatory Archive, KOA, label HI.19990813.45167),
WD~0354$+$463 (HI.19980123.32497), and WD~1455$+$298 (HI.19980624.32907). We supplemented the published data set with
spectra obtained from the KOA for WD~1257$+$278 (HI.20100327.33502, HI.20100327.35353, and HI.20100328.32028)
and WD~1455$+$298 (HI.20060617.24712, HI.20060617.27218, HI.20060617.29698, and HI.20060617.32157).
We used the weighted average of the spectra in our abundance analysis.

A comparative abundance analysis should help us identify
common properties of the sample, or, alternatively, features that are peculiar to NLTT~25792.
The spectral energy distributions (SED) of the comparison stars (see Appendix~1) are well reproduced by 
a single star model or in the case of WD~0354$+$463 by a binary star template.
The SED of WD~1455$+$298 shows a weak IR excess in the WISE bands ($\lambda \gtrsim 3\mu$m) suggesting the presence of warm dust; \citet{far2008}
noted a weak excess in {\it Spitzer} IRAC measurements at $\lambda \gtrsim 5\mu$m and inferred a dust temperature of 400~K.

The prototypical DAZ white dwarf G74-7 \citep[WD~0208$+$396][]{lac1983,bil1997,zuc2003} lies in the Galactic plane ($l=139.2,\,b=-20.4$)
at a distance of $\sim17$ pc \citep[$\pi=0.0598\pm0.0035$][]{van1995} and toward a relatively tenuous
line-of-sight ($E_{\rm B-V} = 0.0485$).

The DAZ white dwarf WD~0354$+$463 \citep{zuc2003} and its dM7 companion form an unresolved,
possibly close pair \citep[sep.$<0.8$ AU,][]{far2006}. \citet{zuc2003} noted emission in the H$\beta$ line core, 
but no radial velocity variations have been reported to date. The presence
of heavy elements in the atmosphere of the white dwarf may be attributed to
a wind-accretion mechanism \citep{zuc2003,deb2006}. The binary lies in the Galactic plane ($l=153.2,\,b=-5.1$) 
and in a relatively dense line-of-sight ($E_{\rm B-V} = 0.642$).
The SED of WD~0354$+$463 is well reproduced by the combination of a DAZ white dwarf 
model and a M7 template constructed using optical/IR spectroscopy of VB~8 \citep[$=$GJ~644~C;][]{tin1998,ray2009,cus2005}.
The star VB~8 is located at a distance of $6.5\pm0.1$ pc \citep{tho2003}.
This exercise demonstrates that the optical/ultraviolet part of the
spectrum is dominated by the DAZ white dwarf. Therefore, the Balmer line analysis is probably correct although weak emission
line cores should be excluded from the analysis. However, the spectral decomposition based on the M7 template implies 
a larger distance ($d=43.2\pm0.7$ pc) than estimated using the white dwarf absolute magnitude ($d=33.4\pm2.5$ pc). 
The distance estimates are reconciled by adopting for the companion an absolute magnitude 
$M_K=10.32$, or a sub-type between M8 and M9 \citep{kir1994} and 0.56 mag fainter than the template itself \citep[$M_K ({\rm VB~8}) = 9.78$ mag,][]{kir1994}. 
Alternatively, the white dwarf itself could be 0.56 mag brighter implying a lower surface gravity than measured spectroscopically.

Finally, both WD~1257$+$278 \citep{zuc2003,zuc2011} and WD~1455$+$298 \citep{zuc2003} are relatively more distant at $\sim35$ pc \citep{van1995},
but at high Galactic latitudes of $b=88.1$ and $62.1$, respectively, and correspondingly low
extinction in the line-of-sight ($E_{B-V} = 0.0095$ and 0.0177, respectively).
No IR calcium triplet emission have been detected in the HIRES spectra of WD~1257$+$278 and WD~1455$+$298.

\begin{figure*}
\epsscale{1.0}
\plotone{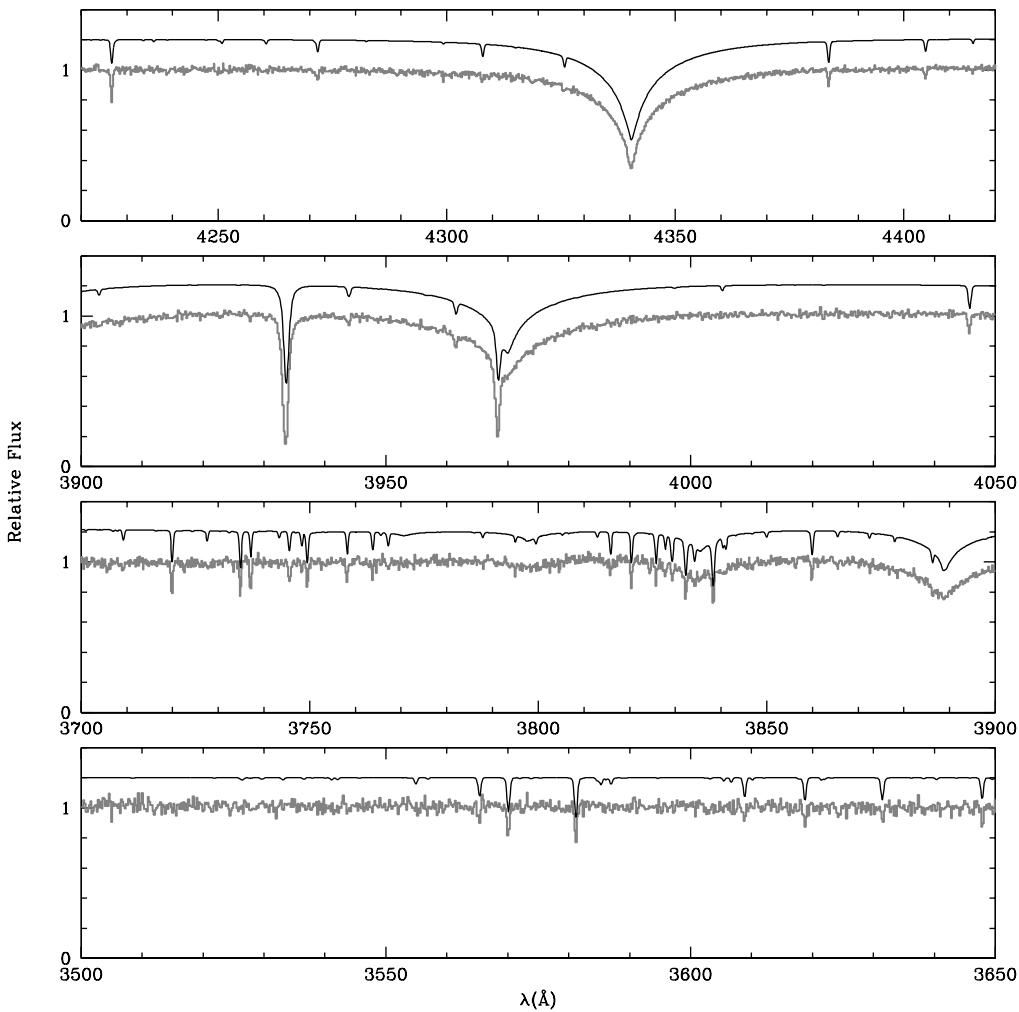}
\caption{Sections of the X-shooter spectra from 3500 to 4420 \AA\ and showing many spectral features listed
in Table~\ref{tbl_line}. The data are compared to a representative model with $[{\rm X/H}]=-2.5$.
\label{fig1}}
\end{figure*}

\section{Analysis}

We based our analysis of the Balmer and heavy element line profiles on a grid of model atmospheres in local thermodynamic
equilibrium that include convective energy transport.  We employed the mixing-length formalism with parameters ML2 and
$\alpha=0.6$. Details of Balmer line profiles are provided by \citet{kaw2006}. Heavy-element contributions to the electron density
are included with the metallicity varying from $[{\rm X/H}]\equiv \log{\rm X/H}-\log{\rm X/H}_\odot=-4.0$ to $-2.0$, where X includes the 18 most abundant
species from carbon to zinc. Based on the model atmosphere structure we computed detailed heavy element line profiles using Voigt functions and state-of-the-art oscillator strengths
and broadening parameters \citep[see details in][]{kaw2012b}; In most cases, collisions with neutral hydrogen atoms, and, to a lesser extent, electrons dominate the line profiles. 

The solar abundance scale employed in the present work was built using the compilations of
\citet{asp2009} and \citet{lod2009}. \citet{lod2009} provide a critical compilation of meteoritic, i.e.,
the CI carbonaceous chondrites, and solar photospheric abundances. Employing somewhat different criteria,
\citet{asp2009} list solar photospheric abundances that differ on average by only 0.005 dex with a dispersion
of 0.04 dex from those of \citet{lod2009} for a group of abundant elements comprising Na, Mg, Al, Si, Ca, and Fe.
Note that for that same group of elements, the CI-chondrites and solar photospheric abundances are nearly
identical with differences no larger than 0.02 dex \citep{lod2009}. 
Therefore, in this work, we use the straight average of the solar photospheric abundances of \citet{asp2009} 
and the CI-chondrites and solar photospheric abundances of \citet{lod2009}.
We will refer to this joint scale as the ``solar abundances'':
$\log{\rm Na/H}_\odot = -5.72$,
$\log{\rm Mg/H}_\odot = -4.44$,
$\log{\rm Al/H}_\odot = -5.54$,
$\log{\rm Si/H}_\odot = -4.48$,
$\log{\rm Ca/H}_\odot = -5.67$,
$\log{\rm Fe/H}_\odot = -4.53$.
Finally, \citet{lod2009} proposes to scale proto-solar (or ``solar system'') abundances from solar abundances using the logarithmic
relation $X_0=X+0.053$. Abundance ratios are not affected by this scaling and the proto-solar abundances 
will not be considered further.

We fitted the Balmer line profiles and extracted the effective temperature and surface gravity (Section 3.2) and constrained the abundance of
individual elements (Section 3.3) using $\chi^2$ minimization techniques.

\subsection{NLTT~25792: Line Identifications and Radial Velocity}

\begin{deluxetable}{lrc}
\tablecaption{Equivalent widths and line velocities\label{tbl_line}}
\tablewidth{0pt}
\tablehead{
\colhead{Ion,$\lambda$ (\AA) \tablenotemark{a}} & \colhead{E.W.(m\AA)} & \colhead{$v$ (km s$^{-1}$) \tablenotemark{b}}  
}
\startdata
\ion{Fe}{1} 3440.606 &  110. &    33.9 \\
\ion{Fe}{1} 3565.379 &   37. &    24.1 \\
\ion{Fe}{1} 3570.097 &   99. &    25.6 \\
\ion{Fe}{1} 3581.193 &   97. &    25.2 \\
\ion{Fe}{1} 3719.935 &  105. &    24.7 \\
\ion{Fe}{1} 3722.563 &   23. &    22.5 \\
\ion{Fe}{1} 3733.317 &   18. &    28.8 \\
\ion{Fe}{1} 3734.864 &  110. &    24.7 \\
\ion{Fe}{1} 3737.131 &  107. &    22.8 \\
\ion{Fe}{1} 3745.561 &   76. &    30.0 \\
\ion{Fe}{1} 3748.262 &   30. &    25.2 \\
\ion{Fe}{1} 3749.485 &   74. &    25.3 \\
\ion{Fe}{1} 3758.233 &   62. &    23.3 \\
\ion{Fe}{1} 3763.789 &   37. &    25.3 \\
\ion{Fe}{1} 3767.191 &   33. &    27.6 \\
\ion{Fe}{1} 3795.002 &   22. &    26.4 \\
\ion{Fe}{1} 3804.791 &   19. &    17.4 \\
\ion{Fe}{1} 3815.840 &   24. &    24.9 \\
\ion{Fe}{1} 3820.425 &   76. &    23.6 \\
\ion{Fe}{1} 3824.444 &   32. &    22.8 \\
\ion{Fe}{1} 3825.881 &   47. &    19.8 \\
\ion{Fe}{1} 3827.823 &   45. &    23.7 \\
\ion{Fe}{1} 3829.452 \tablenotemark{c} &   37. &    20.7 \\
\ion{Mg}{1} 3832.300 &   79. &    22.7 \\
\ion{Fe}{1} 3834.222 &   44. &    28.2 \\
\ion{Mg}{1} 3838.292 &  116. &    24.2 \\
\ion{Fe}{1} 3856.371 &   25. &    26.3 \\
\ion{Fe}{1} 3859.911 &   70. &    23.5 \\
\ion{Ca}{2} 3933.660 & 1215. &    19.6 \\
\ion{Al}{1} 3944.006 &   23. &    21.3 \\
\ion{Al}{1} 3961.520 &   48. &    27.0 \\
\ion{Ca}{2} 3968.470 &   ... &    22.1 \\
\ion{Fe}{1} 4045.812 &   64. &    21.5 \\
\ion{Fe}{1} 4063.594 &   26. &    23.4 \\
\ion{H}{1} 4101.734 &   ... &    17.0 \\
\ion{Ca}{1} 4226.730 &   96. &    26.0 \\
\ion{Fe}{1} 4271.760 &   37. &    22.3 \\
\ion{H}{1} 4340.462 &   ... &    19.0 \\
\ion{Fe}{1} 4383.545 &   46. &    23.9 \\
\ion{Fe}{1} 4404.750 &   32. &    23.3 \\
\ion{Fe}{1} 4415.122 &    9. &    23.5 \\
\ion{H}{1} 4861.323 &   ... &    22.6 \\
\ion{Fe}{1} 5167.488 &  22.  &    33.5 \\
\ion{Mg}{1} 5172.684 &   40. &    27.0 \\
\ion{Mg}{1} 5183.604 &   66. &    25.4 \\
\ion{H}{1} 6562.797 &   ... &    26.9 \\
\ion{Ca}{2} 8498.020 &   58. &    31.7 \\
\ion{Ca}{2} 8542.090 &  273. &    29.5 \\
\ion{Ca}{2} 8662.140 &  158. &    29.2 
\enddata
\tablenotetext{a}{Laboratory wavelength from the server \url{http://www.nist.gov/pml/data/asd.cfm} at the National Institute of Standards and Technology (NIST).}
\tablenotetext{b}{Heliocentric velocities.}
\tablenotetext{c}{Blended with \ion{Mg}{1} $\lambda$3829.3549\AA.}
\end{deluxetable}

\begin{figure*}
\epsscale{1.15}
\plottwo{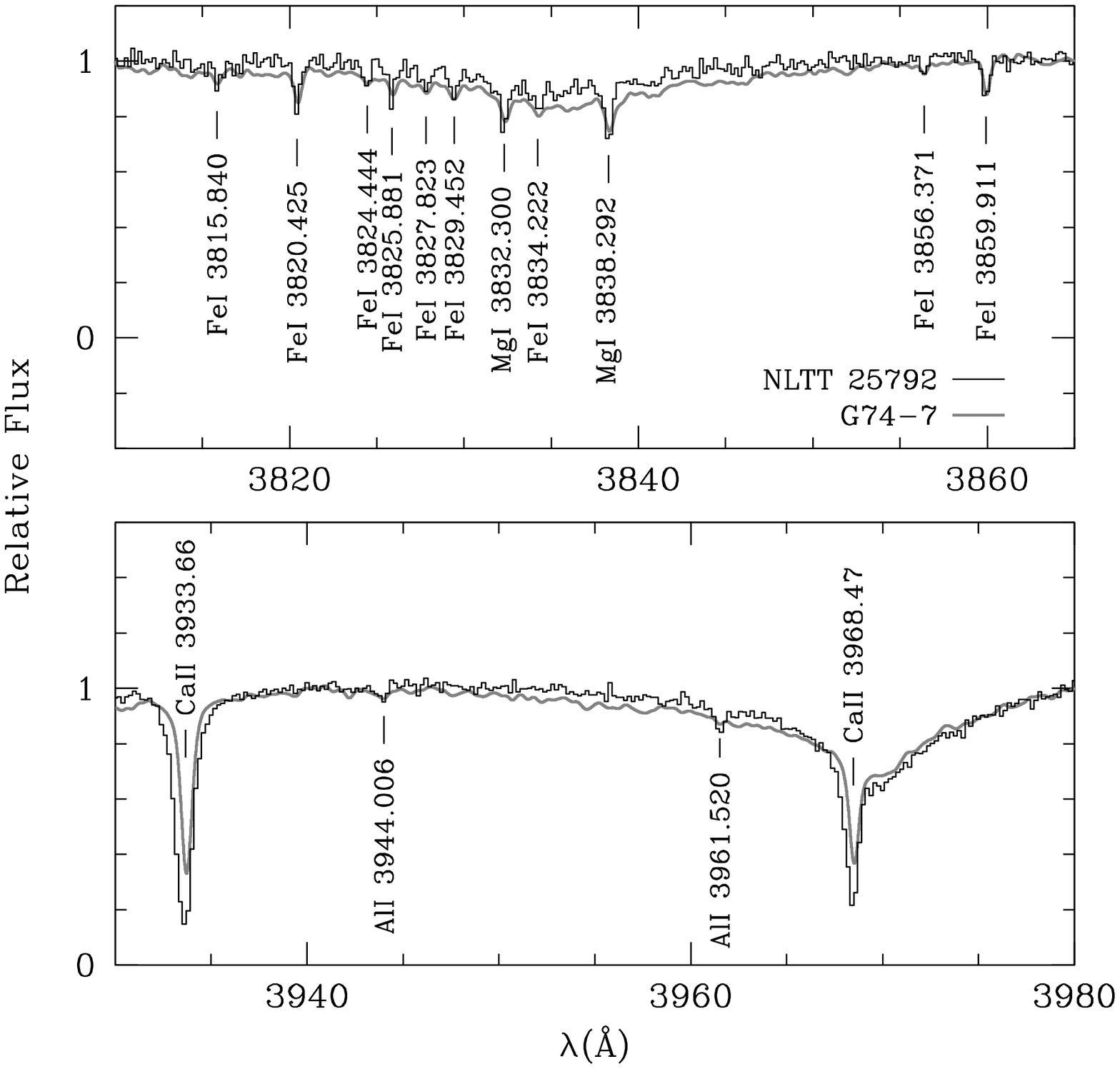}{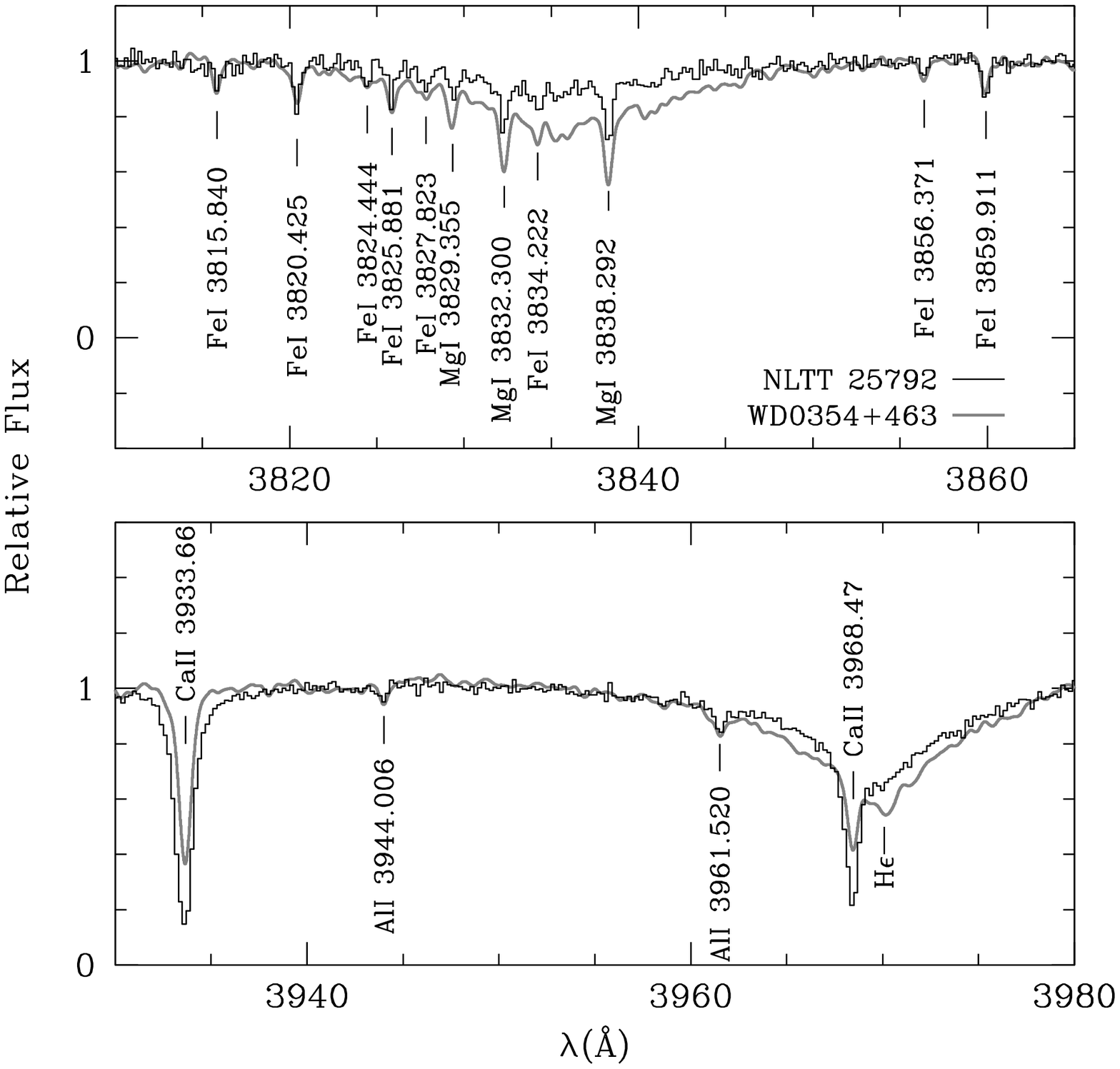}
\caption{Comparing the Ca~H\&K doublet (lower panel) and ultraviolet \ion{Mg}{1}/\ion{Fe}{1} lines (top panel) in the X-shooter spectra of NLTT~25792 (black lines) 
with Keck/HIRES spectra other DAZ white dwarfs (grey lines): (left panels) G74-7 (WD~0208+396) and (right panels) WD~0354+463 (DAZ+dM). The HIRES spectra have been
degraded to the X-shooter resolution. \label{fig2}}
\end{figure*}

Figure~\ref{fig1} shows segments of the UVB X-shooter spectrum. Notable features include the Ca~H\&K doublet and the upper Balmer lines
(H$\gamma$ to H8) and numerous \ion{Fe}{1} lines. 
The average velocity of 49 spectral lines (Table~\ref{tbl_line}) found between $\sim3440$ and $\sim8662$\AA\ is 
24.7 \kms\ with a dispersion of only 3.6 \kms.
Estimating the gravitational redshift of the white dwarf at $31.7\pm1.5$ \kms\ (Section 3.2), the radial velocity of the white dwarf
is $v_r = -7.0\pm3.9$ \kms. The velocity difference between the two consecutive ($\Delta t\approx 1$ hr) exposures is $v_1-v_2 = -2.2$ \kms\
with a dispersion of 4.1 \kms.

Figure~\ref{fig2} compares the main spectral features in NLTT~25792 and the comparison stars. These objects share many important spectral features,
most notably dominant Ca~H\&K doublets and rich iron line spectra. 
A shift of the Ca~K line in NLTT~25792 relative to the other two stars is apparent when lining-up the spectra with other narrow metal lines.
The IR calcium triplet in NLTT~25792 is in absorption with no evidence of an emission component.

\subsection{Atmospheric Parameters and Spectral Energy Distribution (SED) of NLTT~25792}

We fitted the Balmer line profiles, H$\beta$ to H$_{10}$ excluding H$\epsilon$, in the X-shooter and FORS1 \citep{kaw2011b} spectra independently.
Our new measurements, (\teff$,\log{g})=(7900\pm20\,{\rm K},8.09\pm0.04$) with X-shooter and ($7910\pm20\,{\rm K},7.96\pm0.05$) with FORS1, are
in excellent agreement with the measurements of \citet{gia2011}: (\teff$,\log{g})=(7910\pm118\,{\rm K},8.05\pm0.08$). 
We adopted the weighted averages of the measurements: (\teff$,\log{g})= (7903\pm16\,{\rm K},8.04\pm0.03$).
Based on these parameters, the mass of the white dwarf is $0.618\pm0.018\,M_\odot$ and the radius is $0.0124\pm0.0002\,R_\odot$ with
a cooling age of 1.2 Gyr. The absolute magnitude in the SDSS $r$ band, $M_r=13.18\pm0.04$ mag, locates the star at 
a distance of $d=36.2\pm0.7$ pc. 

\begin{figure}
\epsscale{1.15}
\plotone{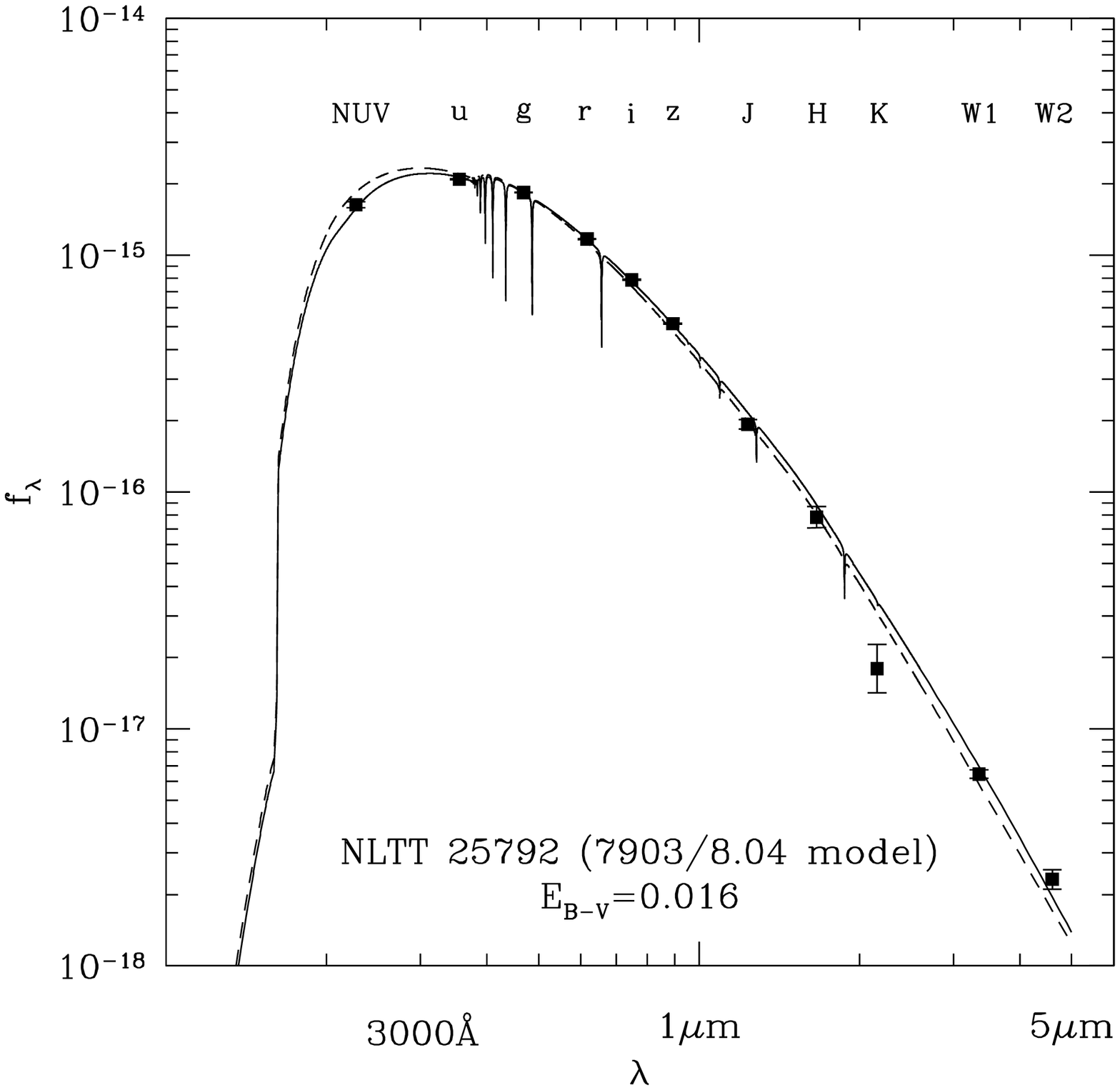}
\caption{Spectral energy distribution, $f_\lambda$ (erg cm$^{-2}$ s$^{-1}$ \AA$^{-1}$) vs $\lambda$, of the DAZ NLTT~25792 from the
near UV to the near-IR (Table~\ref{tbl_phot}). The data are compared to the best-fit model to the Balmer lines including the
effect of interstellar extinction (full line) or excluding it (dashed line). The K-band shows a possible flux deficit.
The {\it GALEX} NUV and SDSS $u$ magnitudes are best fit with the model including interstellar extinction.
\label{fig3}}
\end{figure}

The SED of NLTT~25792 (Fig.~\ref{fig3}) is characteristic of a 7900~K white dwarf without a low-mass stellar companion. Also,
the SED does not show a warm disk often encountered in the IR spectra of polluted white dwarfs \citep{kil2006,far2009}.    
In fact, the measured IR flux shows an unexplained dip measured in the K-band.
However, we noted the possible effect of interstellar reddening on the ultraviolet part of the SED with $E_{B-V} = 0.016$ using the
models of \citet{car1989} and $R_V=3.1$.

\subsection{Heavy Element Abundance Pattern in NLTT~25792}

We measured the abundance of magnesium, aluminum, calcium and iron using spectral lines listed in Table~\ref{tbl_line}. We fitted 
the line profiles rather than the equivalent widths. These are provided as indicative of the relative line strengths.
Following a procedure adopted in the past \citep{kaw2012b}, we employed the broadening parameter $\Gamma$ from
\citet{bar2000} added to Stark and natural broadening parameters. The resulting Voigt profiles are convolved with Gaussian profiles set to match the
spectral resolution.
The measured abundances are listed in Table~\ref{tbl_abun}. We noted a discrepancy between the calcium abundance based
on the Ca~K line (${\rm [Ca/H]} = -2.14\pm0.06$) and that 
based on \ion{Ca}{1} and the \ion{Ca}{2} triplet (${\rm [Ca/H]} = -2.40\pm0.06$). We excluded the Ca~H\&K lines from the abundance measurement
and further investigate the question in Section 3.5.

The upper limit to the equivalent width of \ion{Si}{1}$\lambda$3905.523 line is $\sim10$\,m\AA\ and the upper limits to
the \ion{Na}{1}$\lambda\lambda$5889.951,5895.924 lines are $\sim15$\,m\AA. These estimates correspond to 
$3\sigma$ upper limits, i.e., 
$3\times \Delta\lambda/{\rm SNR}$, where $\Delta\lambda$ is the full width of a resolution element and the
SNR is measured within bins the size of a resolution element. For example, $\Delta\lambda = 0.50$ \AA\ near \ion{Fe}{1}$\lambda$4415 and the measured SNR is
90 per 0.2 \AA\ bin, or 142 per resolution element, corresponding to a minimum equivalent width of $\sim10$ m\AA.
The corresponding silicon and sodium abundance upper limits are $\log{\rm Si/H} \lesssim -7.5$
([Si$/$H]$\lesssim -3.0$) and 
$\log{\rm Na/H} \lesssim -8.8$
([Na$/$H]$\lesssim -3.1$).
Silicon and sodium are markedly depleted relative to all other elements (Mg, Al, Ca, and Fe) on the solar abundance scale.
Also, we noted a possible ISM component to the \ion{Na}{1}$\lambda$5889.951 line at $-13$\,\kms\ and with an equivalent width of 35\,m\AA.
We did not attempt to constrain the CNO abundance because of a lack of practical abundance diagnostics at a temperature of $\sim7900$ K for NLTT~25792. 
For example, the \ion{O}{1}$\lambda$7773 triplet remains extremely weak under these conditions.

The systematic effects of effective temperature and surface gravity variations have been investigated.
The abundance shifts corresponding to surface gravity shifts of $\pm0.1$ dex
for a reference model at \teff$=7900$~K and $\log{g}=8.0$ are negligible when measuring calcium or aluminum abundances, but amount to
$\mp 0.01-0.02$ in the logarithm of the magnesium abundance and $\pm0.02-0.04$ for iron.
The effect of temperature shifts of $\pm100$~K 
for the same reference model are $\pm0.06$ for calcium and aluminum, and $\pm0.04$ for magnesium and iron.
In summary, the temperature uncertainty dominates the errors in abundances relative to hydrogen, but because the relevant elements
follow the same trends it would not affect abundance ratios. 

\subsection{Comparative Analysis}

\begin{deluxetable*}{ccccccc}
\tablecaption{Properties and abundances\label{tbl_abun}}
\tablewidth{0pt}
\tablehead{
\colhead{} & \colhead{NLTT~25792} \tablenotemark{a} & \colhead{G74-7} & \colhead{WD~1455+298} & \colhead{WD~0354+463} & \multicolumn{2}{c}{WD~1257+278 \tablenotemark{a}}}
\startdata
\teff (K)                         &  $7903\pm16$     & $7306\pm22$      & $7383\pm19$      & $8240\pm120$     & $8609\pm20$      & (8600)   \\
$\log{g}({\rm cm\,s^{-2}})$       &  $8.04\pm0.03$   & $8.06\pm0.02$    & $7.97\pm0.03$    & $7.96\pm0.10$    & $8.24\pm0.02$    & (8.0)    \\
 & & & & & \\
$\log{\rm Mg/H}$                  & $-$7.24$\pm$0.05 & $-$7.79$\pm$0.06 & $-$8.03$\pm$0.06 & $-$6.70$\pm$0.05 & $-$7.49$\pm$0.08 & $-$7.51$\pm$0.09 \\
${\rm [Mg/H]}$ \tablenotemark{b}  & $-$2.80$\pm$0.05 & $-$3.35$\pm$0.06 & $-$3.59$\pm$0.06 & $-$2.26$\pm$0.05 & $-$3.05$\pm$0.08 & $-$3.07$\pm$0.09 \\
 & & & & & \\
$\log{\rm Al/H}$                  & $-$8.16$\pm$0.11 & $-$8.90$\pm$0.20 &       ...        & $-$7.98$\pm$0.13 & $-$8.50$\pm$0.25 & $-$8.50$\pm$0.25 \\
${\rm [Al/H]}$ \tablenotemark{b}  & $-$2.62$\pm$0.11 & $-$3.36$\pm$0.20 &       ...        & $-$2.44$\pm$0.13 & $-$2.96$\pm$0.25 & $-$2.96$\pm$0.25 \\
 & & & & & \\
$\log{\rm Ca/H}$                  & $-$8.07$\pm$0.06 & $-$9.05$\pm$0.04 & $-$9.51$\pm$0.03 & $-$8.20$\pm$0.03 & $-$8.38$\pm$0.06 & $-$8.39$\pm$0.06 \\
${\rm [Ca/H]}$ \tablenotemark{b}  & $-$2.40$\pm$0.06 & $-$3.38$\pm$0.04 & $-$3.84$\pm$0.03 & $-$2.53$\pm$0.03 & $-$2.71$\pm$0.06 & $-$2.72$\pm$0.06  \\
 & & & & & \\
$\log{\rm Fe/H}$                  & $-$7.16$\pm$0.04 & $-$8.03$\pm$0.09 & $-$8.40$\pm$0.08 & $-$7.13$\pm$0.11 & $-$7.47$\pm$0.09 & $-$7.45$\pm$0.10 \\
${\rm [Fe/H]}$ \tablenotemark{b}  & $-$2.63$\pm$0.04 & $-$3.50$\pm$0.09 & $-$3.87$\pm$0.08 & $-$2.60$\pm$0.11 & $-$2.94$\pm$0.09 & $-$2.92$\pm$0.10 
\enddata
\tablenotetext{a}{The calcium abundance measurement excludes Ca H\&K.}
\tablenotetext{b}{$[{\rm X/H}] = \log{\rm X/H}-\log{\rm X/H}_\odot$.}
\end{deluxetable*}

The abundance analysis of the four related stars is performed for given atmospheric parameters (Table~\ref{tbl_abun}).
We collected published effective temperature and surface gravity
measurements based on Balmer line profile analyses or joint line profile and parallax analyses.

For G74-7 we averaged the temperature and gravity measurements of \citet{bil1997}, \citet{gia2004}, \citet{gia2005}, \cite{hol2008},
\citet{gia2011}, and \citet{gia2012} that are based on a Balmer line profile analysis (method 1), and compared the results to
the average of the measurements of \citet{ber1997}, \citet{leg1998}, and \citet{ber2001} that
are based on the parallax of \citet{van1995}, optical/IR SEDs, and high-dispersion H$\alpha$ spectroscopy (method 2). 
Some of these measurements may well be redundant, but, in general, 
they should reflect on differing data sources or model atmosphere generations.
In this case, the two methods delivered consistent results: (\teff, $\log{g})=7305\pm22,\,8.07\pm0.03)$ with method 1
and (\teff, $\log{g})=7316\pm103,\,8.02\pm0.05)$ with method 2. Therefore, we adopted the weighted average of all
temperature and surface gravity measurements (Table~\ref{tbl_abun}). 
The corresponding abundance measurements differ slightly from published values: we obtain
a lower calcium abundance ($-0.2$ dex)
but a higher magnesium abundance (0.3 and 0.1 dex) than in \citet{bil1997} and \citet{zuc2003}, although our iron and aluminum abundance
measurements are in good agreement (within $\approx 0.1$ dex) with those of \citet{zuc2003}.
The adopted atmospheric parameters in either study are similar to those adopted in this work. Variations may be
attributed to different fitting techniques or model generations.
The pattern is very nearly scaled on the solar pattern with $[{\rm X/H}]\approx -3.4$, where X represents Mg, Al, Ca, and Fe.

Next, for WD~0354+463 we used the Balmer line analysis of \citet{gia2011}. The effective temperature adopted by \citet{zuc2003} was
close to 500~K cooler and should affect abundance measurements. Predictably, because of the higher temperature adopted 
in this work our abundance measurements are on average 0.19 dex higher with a dispersion of only 0.08 dex. However, the patterns are 
similar, except for a slight magnesium enrichment in our data.
In addition, and as in the case of G74-7, the calcium abundance measurements based on \ion{Ca}{1}$\lambda$4226 and Ca~K are formally
in agreement. 

For WD~1257$+$278 we averaged the measurements of \citet{gia2011}, \citet{lim2010}, \citet{hol2008}, and \citet{lie2005}
to which we compounded our own analysis of two available SDSS spectra. We excluded a notably defective H$\alpha$ line from 
the analysis of one of the SDSS spectrum. 
We obtained similar parameters (Table~\ref{tbl_abun}) to those adopted by \citet{zuc2011}.
The surface gravity (hence mass) obtained in the joint astrometric/photometric/spectroscopic analysis of  \citet{ber2001}
is lower than estimated in the cited spectroscopic analyses. Comparing the last two columns of Table~\ref{tbl_abun} we conclude
that the effect on the abundance analysis 
of the surface gravity uncertainty is negligible.
However, \citet{zuc2011} list the following abundances for WD~1257$+$278:
$[{\rm Mg/H}] = -2.80$, $[{\rm Al/H}] = -2.63$, $[{\rm Ca/H}] = -2.37$, and $[{\rm Fe/H}] = -2.88$.
Their adopted stellar parameters are nearly identical to ours (\teff$=8600\pm100$ K, $\log{g}=8.10\pm0.15$).
Apart from similar iron abundances, the abundances of Mg, Al, and Ca are approximately
a factor of two lower in our work although we employed the same data set. 
Such discrepancies may, in part, be caused by differences in the model atmospheres or by
different line measurement techniques. The equivalent width measurement of weak lines are notably affected by
the choice of the integration window that may inadvertently include
neighboring lines or uncalibrated continuum variations. Abundance measurements based on a few lines may suffer from
such systematic effects that, however, would tend to average out when including many spectral lines in the abundance measurement.
Spectral line fitting may still suffer from poor continuum placement but the line integration is necessarily
confined to the width of the synthetic line profile. We noted that the present iron abundance measurement and that
of \citet{zuc2011} are based on numerous lines, possibly averaging out systematic effects, and are formally in agreement. 
Overall, and in agreement with \citet{zuc2011}, we found that the atmosphere of WD~1257$+$278 appears relatively rich in calcium, particularly
relative to magnesium.
The calcium abundance measured from the Ca~K line, $[{\rm Ca/H}] = -2.53\pm0.04$, is
0.16 dex higher than measured using the \ion{Ca}{1}$\lambda$4226 line and the Ca~K line
core is poorly fitted, although the overall line profile is well matched. This slight abundance discrepancy
may suggest an effect similar to that observed in NLTT~25792 although the Ca~K line in WD~1257$+$278 does
not show a notable blue shift or asymmetry.

Finally, we averaged the effective temperature and surface gravity measurements of \citet{gia2011}, \citet{hol2008}, \citet{lie2005}, and our own measurement 
of WD~1455$+$298 based
on SDSS spectroscopy. These measurements based on Balmer line profiles are only marginally consistent with the parallax that implies
a larger stellar radius hence lower gravity:
WD~1455$+$298 is a suspected double degenerate \citep{ber2001}. 
However, H$\beta$ and other lines only show single components. Moreover, the SED (Appendix 1) does not show evidence of a cool
companion. This discrepancy remains unresolved. The atmosphere of WD~1455$+$298 is the cleanest of the sample with an 
average metallicity index of only $[{\rm X/H}]=-3.7$ where X represents Mg, Ca, and Fe.
Our new abundance measurements differ on average with those of \citet{zuc2003} by only $-0.1$ dex but with a dispersion of 0.2 dex. 
The increased signal to noise ratio in recent KOA data resulted in more accurate abundance measurements and a clear detection of the \ion{Ca}{1}$\lambda4226$ line. 
The present analysis indicates a modest enrichment in magnesium relative to calcium and iron.
The present calcium and iron abundance measurements in WD~1455$+$298 supercede those presented earlier in \citet{kaw2011b}. 

\begin{figure}
\epsscale{1.15}
\plotone{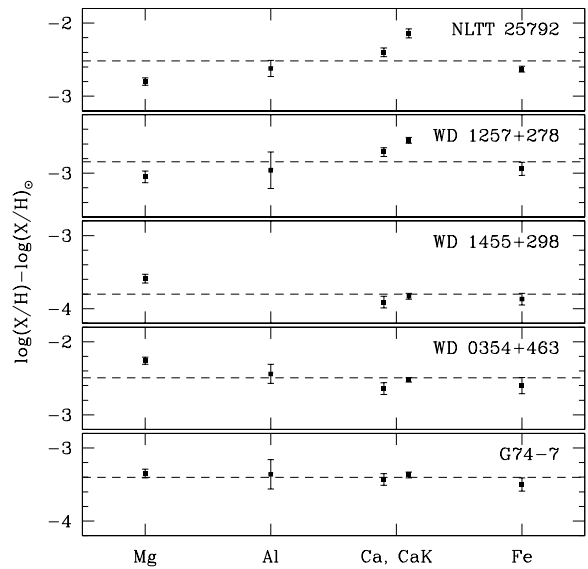}
\caption{Comparative analysis of magnesium, aluminum, calcium, and iron abundances in the five selected stars.
The Ca abundance refers to abundance indicators other than Ca~H\&K, i.e, \ion{Ca}{1}$\lambda$4226 or the \ion{Ca}{2} IR triplet.
The pattern observed in NLTT~25792 is similar to that of WD~1257$+$278 and notably dissimilar to the other patterns (see Section 3.4).
\label{fig4}}
\end{figure}

Figure~\ref{fig4} shows the abundance patterns in the five objects analyzed. The patterns for both G74-7, WD~0354$+$463, and WD~1455$+$298 do not suggest
calcium enhancement, but those of NLTT~25792 and WD~1257$+$278 show a clear enhancement relative to all other elements.
The calcium to magnesium ratio is the most revealing with an enhancement relative to solar
of $+0.40$ in NLTT~25792 and $+0.34$ in WD~1257$+$278.
Interestingly, the patterns for WD~0354+463 and WD~1455+298 show a reversed trend with calcium at its lowest 
abundance relative to magnesium in the sample, $\approx -0.27$ and $-0.25$ below solar. Overall, the abundance pattern in G74-7 is flat, with
a slightly lower iron abundance than the average pattern. Diffusion effects, i.e., the effects of varying diffusion
time scales on the observed abundances, are likely to alter the observed abundance pattern relative to the supplied, i.e, accreted 
pattern (see a discussion in Section 4).

\subsection{A \ion{Ca}{2} Nebular Component?}

\begin{figure}
\epsscale{1.15}
\plotone{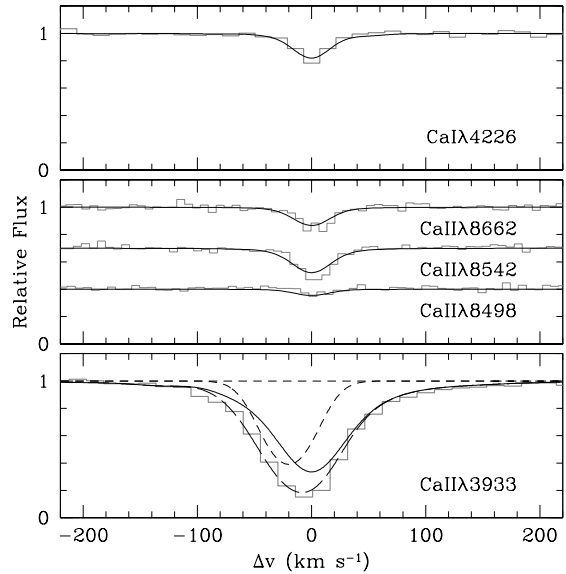}
\caption{The calcium lines in the X-shooter spectra (grey lines) are compared to spectral line syntheses (full lines) with $[{\rm Ca/H}] = -2.4$ (Table~\ref{tbl_abun}). 
The Ca~K line is stronger than
predicted by the model and shows a possible nebular component shifted by $-20$ \kms\ (short dashed line) that we added to the total profile (long dashed line). 
\label{fig5}}
\end{figure}

Modelling of the Ca~K line remains unsatisfactory: The line profile appears deeper and blue-shifted
relative to the best fit model to all calcium lines (Fig.~\ref{fig5}). Incidentally,
the line did not vary in strength between the FORS1 and X-shooter observations.

The line width is dominated by collisions with hydrogen atoms. At the temperature and
density conditions prevalent in the atmosphere of NLTT~25792, we estimate that 80\%
of the total width $\Gamma_{\rm tot}$ is contributed by hydrogen atoms and only 20\% by electrons.
Increasing $\Gamma_{\rm tot}$ by a factor of two does increase the equivalent width
by 33\% although it does not displace the line further toward the blue. On the other hand,
strong lines, such as the Balmer or the Ca~H\&K lines, are visibly affected by the atmospheric 
structure over a wide range of depths, but we noted that the
H$\alpha$ and H$\beta$ line wings and deep cores measured with X-shooter are
well modeled. An accumulation of calcium above the convection zone could result in
a stronger line core than predicted by our homogeneous models.

Alternatively, the excess absorption in the Ca~K line and the velocity offset could be 
interpreted as evidence of a nebular component to the observed profile
with an equivalent width of $\approx 480\,$m\AA\ and at a relative velocity of $-20$ \kms. Correcting for the gravitational redshift ($\sim30$ \kms), the
average velocity of the gas relative to the surface is $+10$ \kms.
Adopting velocity dispersion appropriate for local interstellar gas at $T\sim7000$\,K, i.e., $\sigma=1.7$\,\kms,
an exceedingly large column density $\log{N}$(\ion{Ca}{2}\,cm$^{-2})\gtrsim 15$ would be required to fill in the observed absorption at
a relative velocity of $-20$ \kms. However, allowing a larger velocity dispersion of $\sigma=30$\,\kms, i.e., typical of the range of projected orbital velocity for gas
transiting in front of the stellar disk \citep[see][]{deb2012}, a lower density of $\log{N}$(\ion{Ca}{2}\,cm$^{-2})\approx 12.8$ is found. 
This simple geometrical effect helps locate the gas at a radius $r = (R_{wd}/\sigma)^{2/3} \,(G\,M_{wd})^{1/3}\approx 20\,R_{wd}$ well
within a tidal disruption radius of 100\,$R_{wd}$.
Note that in this analysis it would be more appropriate to adopt rotational broadening function rather than Gaussian velocity distribution.

An origin in the interstellar medium is unlikely.
The absorption
largely exceeds measurements of interstellar \ion{Ca}{2} K line widths ($<300$ m\AA) up
to distances of 400 pc \citep{wel2010}.
Assuming a maximum \ion{Ca}{2} volume density of $10^{-8}$ cm$^{-3}$, the total
column density at the distance of NLTT~25792 ($d\approx 36$ pc) would be $10^{12}$ ions cm$^{-2}$.
In conditions typical of the local ISM, the corresponding line equivalent width would not
exceed 50 m\AA. At a lower, typical volume density of $10^{-9}$ cm$^{-3}$ the column density
is $10^{11}$ cm$^{-2}$ and the corresponding equivalent width is only a few m\AA.
It is therefore unlikely that the excess \ion{Ca}{2} absorption would originate in the
interstellar medium. We conclude that it probably originates in a gaseous circumstellar environment.
The presence of ionized gas in the circumstellar environment of DAZ white dwarfs is well documented \citep[see, e.g.,][]{gae2006,mel2012,deb2012} and revealed mostly by
IR calcium triplet emission, although \citet{gae2012} noted excess \ion{Si}{4} absorption in ultraviolet spectra of
the hot DAZ PG~0843$+$516 that could also originate in a hot circumstellar environment.  
The X-shooter spectra of NLTT~25792 show the IR calcium triplet in absorption. 
The DAZ WD~1257$+$278 also shows excess absorption in the Ca~H\&K lines although the HIRES spectrum does not show
an obvious line shift or asymmetry. Again, the absorption profile resulting from transiting gas follows a
rotational broadening function that may be hidden within the combined profile.

The presence of circumstellar Ca~K line in DAZ spectra is analogous to \ion{C}{4} absorption in the circumstellar environment of hot white dwarfs.
Ionized species of carbon and silicon are found in circumstellar environment of hot white dwarfs.
\citet{ban2003} ascribed those features to a Str\"omgren sphere in the interstellar medium excited by
ultraviolet radiation emanated by the white dwarf, although \citet{dic2012} also cite the possibility
of evaporating circumstellar debris in the intense ultraviolet radiation field. Clearly, young white dwarfs
may be surrounded by even denser material than their older, DAZ counterparts.

\section{Summary and Discussion}

We measured the metallicity in the atmosphere of the DAZ white dwarf NLTT~25792 based on the detection
of magnesium, aluminum, calcium and iron lines in X-shooter spectra of this object.
The average abundance of these elements relative to solar on a logarithmic scale is $[{\rm X/H}]\approx -2.5$ dex.
On the same scale, the upper limits to the sodium and silicon abundances are $[{\rm Na/H}]\lesssim -3.1$ and $[{\rm Si/H}]\lesssim -3.0$.
The atmosphere of NLTT~25792 appears relatively rich in calcium, but relatively poorer in sodium, magnesium, and silicon.
Also, the Ca~K line appeared both deeper and offset than predicted by our models suggesting the presence of circumstellar
gas. The absence of IR calcium triplet emission implies a lack of ionizing radiation concordant with the relatively low
effective temperature of NLTT~25792 (\teff $\approx 7900$\,K) compared to other DAZ white dwarfs with gaseous 
disks \citep[\teff $\gtrsim 20,000$\,K, see, e.g.,][]{gae2006}, or their He-rich counterparts with comparable
ionizing flux \citep[e.g., SDSS~J0738+1835 with \teff$\approx14,000$\, K,][]{duf2012}.
Disk modeling by \citet{har2011} shows that emission lines, such as the IR calcium triplet, 
occurs within the gaseous disk at temperatures of $T_{\rm disk}\lesssim 7000$\,K
and inside the tidal radius.

Chemical diversity in the atmosphere of accreting white dwarfs may be attributed to diversity in source compositions.
For example, the DAZ NLTT~43806 \citep{kaw2006} appears to be
iron deficient prompting \citet{zuc2011} to propose that this deficiency along with a calcium/aluminum
enrichment implies that the white dwarf may
be accreting predominantly ``earth-type lithosphere" material.
Moreover, \citet{gae2012} found evidence of chemical diversity in a sample of
warm ($\sim 20\,000$ K) white dwarfs, particularly in the form of an
overabundance of iron that implies differentiation in the accretion flow. 

Another explanation for chemical diversity is the time dependence of the accretion flow and atmospheric diffusion.
Diffusion time-scales at the bottom of the convection zone \citep[see][]{koe2009} 
\footnote{See also updated time-scale calculations on-line at {\tt http://www1.astrophysik.uni-kiel.de/$\sim$koester/astrophysics/} and dated January 2013.}
range from $\tau\sim 10^3$ to $10^4$ yrs for a hydrogen-rich white dwarf with an effective temperature close to 8000\,K such as NLTT~25792.
This time-scale is nearly instantaneous relative to the age of the white dwarf ($t_{\rm cool} \approx 10^9$ yrs). However, we may well speculate that a time
scale of $10^3$ yrs is comparable or longer than the time lapse between single accretion events that would result in time-variable
abundances. 

This complicated history may be summarized with two extreme examples related to calcium and iron: in the first, our observations took place shortly
($t<<\tau$) after the accretion event and the abundances simply reflect the accretion source. In this case, the body accreted onto NLTT~25792 
must have been moderately enriched in calcium relative to iron.
In the second example, the observations took place long after the accretion event ($t>\tau$) and, following
\citet{koe2009}, the abundances follow an exponential decline $X/H \propto e^{-t/\tau}$, so that the 
abundance ratio of element X relative to Y follows:
\begin{displaymath}
\log{\rm X/Y}-\log{\rm X/Y}_{\rm source} = -\frac{t}{\ln{10}}\,\Big{(}\frac{1}{\tau_{\rm X}} - \frac{1}{\tau_{\rm Y}} \Big{)}.
\end{displaymath}
If $\tau_{\rm X} > \tau_{\rm Y}$, as in the case of calcium (X) versus iron (Y), then X (calcium) would gradually dominate
over Y (iron) on a time-scale comparable to diffusion time-scales. Numerically, $\tau_{\rm Ca}/\tau_{\rm Fe}\approx 1.3-1.5$ in
conditions appropriate for NLTT~25792, so that before all elements disappear from the atmosphere at, say, $t\approx\tau_{\rm Ca}$ 
the calcium abundance would be enhanced by a factor of 1.4 to 1.6 relative to initial conditions.  Assuming initially solar abundances 
this calcium-to-iron abundance ratio is nearly the ratio observed in NLTT~25792 ($\approx 1.5$, see Table~\ref{tbl_abun}) and there would be no need to assume
a calcium-rich accretion source. A similar exercise involving the calcium to magnesium abundance ratio necessarily implies a shortage of magnesium in the
accretion source because their respective diffusion time-scales are nearly equal. The same situation holds for sodium.

In steady state accretion, the observed abundance ratios are simply given by
\begin{displaymath}
\frac{\rm X}{\rm Y} = \frac{\rm X}{\rm Y}_{\rm source} \,\frac{\tau_{\rm X}}{\tau_{\rm Y}}
\end{displaymath}
In this case we would have expected an excess of 0.11 to 0.17 dex of the calcium to iron ratio relative to a source assumed
to be solar, i.e., close to excess observed in NLTT~25792 ($0.23\pm0.10$).
However, examining other abundance ratios should help constrain the abundance pattern of the accretion source.
For example, sodium, magnesium, and calcium have nearly identical diffusion time scales but we found
a significant deficit in sodium and magnesium relative to solar with $[{\rm Na/Ca}] \lesssim -0.7$ and $[\rm Mg/Ca] = -0.4$.
Diffusion time scales for silicon are uncertain. The calculations of \cite{koe2009} indicate a longer time scale for silicon
than calcium, although the recent on-line data indicate the opposite. In either case, the silicon time scale is longer
than that of iron for conditions found in NLTT~25792 implying that the silicon deficit ($[{\rm Si/Fe}] \lesssim -0.4$, $[{\rm Si/Ca}] \lesssim -0.6$) can only be explained by its
absence in the accretion source. 
Assuming steady state, we conclude that the accretion source shows a deficit 
in sodium, magnesium and silicon relative to calcium and iron. The aluminum abundance does not significantly depart from
solar abundance ratios.
Interestingly, \citet{koe2011} finds that a deficit in sodium relative to solar in some DZ stars would occur while accreting ``bulk Earth'' material, although
an explanation for a similar deficit in silicon is not forthcoming.

In summary, and
following \citet{koe2009}, the estimated steady-state accretion rates onto the white dwarf NLTT~25792 are $\dot{M}_{\rm Mg} = 4.6\times10^7$\,g\,s$^{-1}$,
$\dot{M}_{\rm Ca} = 1.2\times10^7$\,g\,s$^{-1}$, and $\dot{M}_{\rm Fe} = 1.7\times10^8$\,g\,s$^{-1}$, and
using on-line data from D. Koester we estimate $\dot{M}_{\rm Al} = 5.3\times10^6$\,g\,s$^{-1}$.

This limited DAZ sample already suggests that the circumstellar environment varies significantly. Abundance ratios, more particularly
$[\rm Mg/Ca]$, vary considerably within this sample, from $[\rm Mg/Ca] = -0.4$ in NLTT~25792
to $[\rm Mg/Ca] = +0.27$ in WD~0354$+$463. The reasons for these variations are not known.
The ratios $[{\rm Na/Ca}]$ and $[{\rm Si/Ca}]$ in NLTT~25792 fall well
below solar ratios and this deficit must originate in the accretion source, although
the precise circumstances surrounding accretion of circumstellar material onto cool white dwarfs remain uncertain.
In this context, a correlation found between condensation temperature of accreted constituents and corresponding
photospheric abundances in a helium-rich polluted white dwarf (DBZ) may offer come clues to the exact nature of
the accretion mechanism \citep{duf2012}.
In a broader context, the calcium to iron abundance ratio in DZ and DAZ stars alike is known to vary by well over
an order of magnitude \citep[see][]{jur2013} although we find this ratio to be rather homogeneous within our sample. 
On the other hand, \citet{koe2011} measured large dispersions (0.4 to 0.6 dex) in the abundance ratio distributions
(Na, Mg, Ca, and Fe)
of a large DZ sample.
The present study shows that such variations occur between magnesium and other elements within a sample of
closely related DAZ white dwarfs.

New spectroscopy with a higher dispersion than achieved with X-shooter ($R\approx 9000$) would be useful in
resolving the Ca~K line profile into its photospheric and circumstellar components, if any.
More importantly,
the existence of polluted, magnetic white dwarfs
\citep[e.g., NLTT~10480 and NLTT43806;][]{kaw2011,zuc2011} may be linked to field-generating
accretion or interaction events \citep{tou2008,nor2011}. The detection of magnetic fields weaker than 20 kG is
not possible at the resolution achieved with X-shooter but high-dispersion spectroscopy of cool
DAZ white dwarfs may help reinforce a link between accretion events and magnetic field generation in
compact objects.

\acknowledgments

A.K. and S.V. acknowledge support from the Grant Agency of the Czech Republic
(P209/12/0217 and 13-14581S). This work was also supported by the project
RVO:67985815 in the Czech Republic. We thank the referee for suggesting several
improvements to the paper.

This research has made use of the Keck Observatory Archive (KOA), which is operated 
by the W. M. Keck Observatory and the NASA Exoplanet Science Institute (NExScI), 
under contract with the National Aeronautics and Space Administration.

This publication makes use of data products from the Wide-field Infrared Survey Explorer, which 
is a joint project of the University of California, Los Angeles, and the Jet Propulsion 
Laboratory/California Institute of Technology, funded by the National Aeronautics and Space Administration, 
and from the Two Micron All Sky Survey, which is a joint project of 
the University of Massachusetts and the Infrared Processing and Analysis Center/California Institute of 
Technology, funded by the National Aeronautics and Space Administration and the National Science Foundation. 

{\it Facilities:} \facility{VLT:Kueyen}

\appendix

\begin{figure*}
\epsscale{1.15}
\plotone{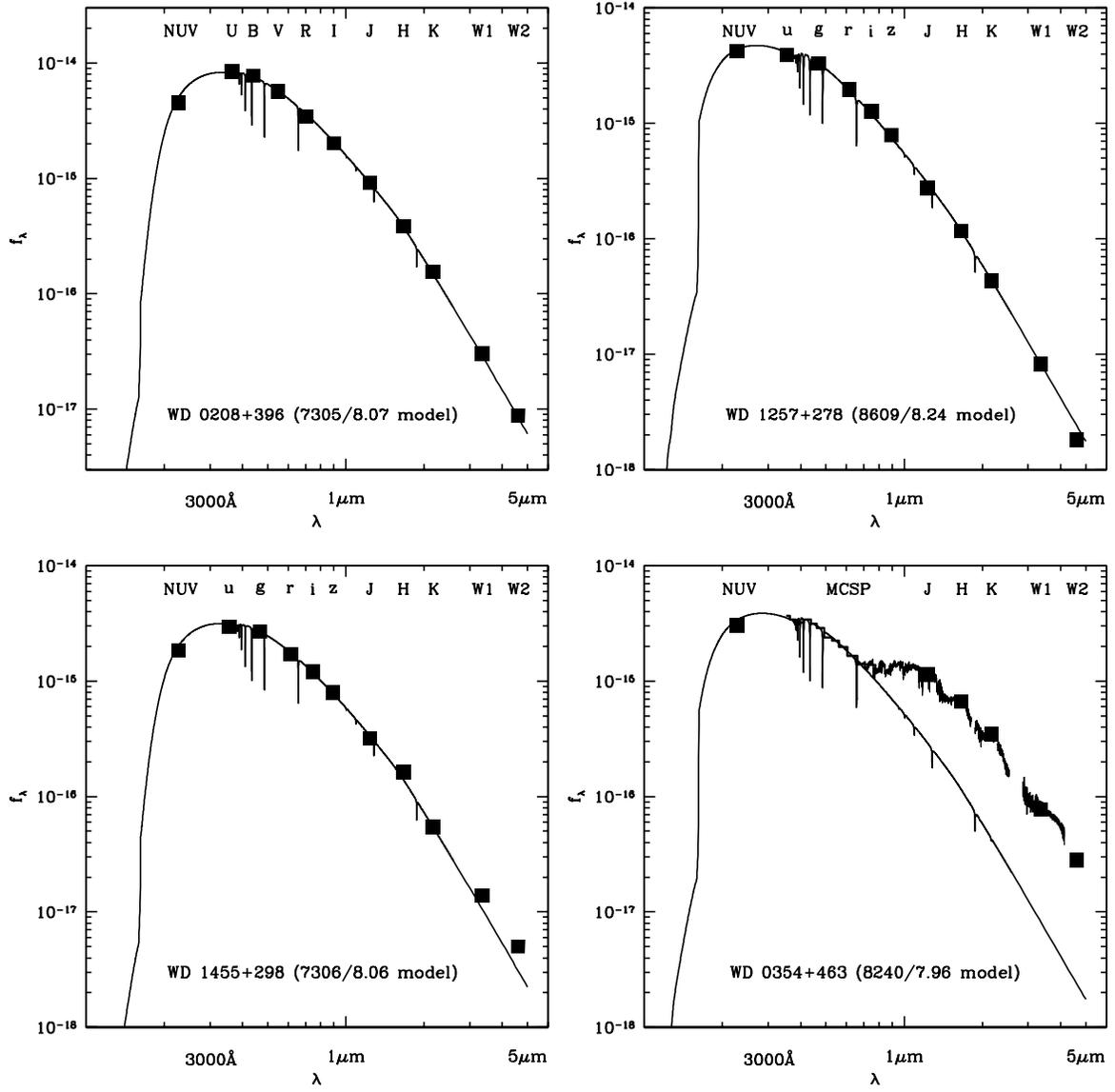}
\caption{Spectral energy distributions of the four related stars. The excess in WD~0354$+$463 is due to the dM7 companion, while a weak near-IR
excess is apparent in the {\it WISE} observations of WD~1455$+$298.
\label{fig6}}
\end{figure*}

The SEDs of the four related stars (Fig.~\ref{fig6}) reveal two apparently single white dwarfs without clear evidence of
circumstellar environments and one single white dwarf with a faint dusty environment \citep[WD~1455$+$298; ][]{far2008}, while the SED of WD~0354$+$463 shows a dM7 companion. The SEDs were built using {\it GALEX} NUV, SDSS ugriz, 2MASS JHK, and {\it WISE} photometric measurements (see Section 2). Optical data for WD~0354$+$463 are from \citet{gre1990}
using the multichannel spectrophotometer (MCSP), and from \citet{hol2008} for WD~0208$+$396.

The data are compared to predicted SED based on adopted stellar parameters (Table~\ref{tbl_abun}). 

\end{document}